\input{aipcheck}

\documentclass[
    ,final            % use final for the camera ready runs
  ]
  {aipproc}

\layoutstyle{6x9}
\def\eslt{E_T^{\rm miss}}

\def\to{\rightarrow}

\def\bi{\begin{itemize}}
 \def\ei{\end{itemize}}

\def\c1p{C1^\prime}
\def\ta{\tilde a}

\def\ta{\tilde a}

\def\tg{\tilde g}

\def\tw{\tilde\chi^\pm}

\def\tz{\tilde\chi^0}
\def\alt{\stackrel{<}{\sim}}
\def\agt{\stackrel{>}{\sim}}
\def\be{\begin{equation}}  
\def\ee{\end{equation}}  
\def\bea{\begin{eqnarray}}  
\def\eea{\end{eqnarray}}

\newcommand\sjp[3]{{\it Sov.\ J.\ Nucl.\ }{\bf #1} (#2) #3}
\newcommand\prd[3]{{\it Phys.\ Rev.\ }{\bf D #1} (#2) #3}

\newcommand\prl[3]{{\it Phys.\ Rev.\ Lett.\ }{\bf #1} (#2) #3}
\newcommand\plb[3]{{\it Phys.\ Lett.\ }{\bf B #1} (#2) #3}
\newcommand\jhep[3]{{\it J. High Energy Phys.\ }{\bf #1} (#2) #3}

\newcommand\npb[3]{{\it Nucl.\ Phys.\ }{\bf B #1} (#2) #3}

\newcommand\arnps[3]{{\it Ann.\ Rev.\ Nucl.\ Part.\ Sci.}{\bf  #1} (#2) #3}

\begin{document}

\title{$SO(10)$ SUSY GUTs with \\
mainly axion cold dark matter:\\
implications for cosmology and colliders}


\author{Howard Baer
\footnote{Invited talk given at Axions 2010 conference, 
January 15-17, 2010, University of Florida, Gainesville, FL USA.}
}{
  address={Dep't of Physics and Astronomy, University of Oklahoma, Norman, OK, 73019, USA}
}

\begin{abstract}
Supersymmetric grand unified theories based on the gauge group
$SO(10)$ are highly motivated. In the simplest models, one expects 
$t-b-\tau$ Yukawa coupling unification, in addition to gauge, matter 
and Higgs unification. Yukawa unification only occurs with very special
GUT scale boundary conditions, leading to a spectra with $\sim 10$ TeV
first and second generation scalars, TeV-scale third generation scalars, and
light gauginos. The relic density of neutralino cold dark matter 
is calculated to be $10^2-10^4$ times higher than observation. If we
extend the theory with the PQ solution to the strong CP problem, then
instead a mixture of axions and axinos comprises the dark matter,
with the measured abundance.
Such a solution solves several cosmological problems. We predict a 
rather light gluino with $m_{\tg}\sim 300-500$ GeV that should be
visible in either Tevatron or forthcoming LHC run 1 data. We would also
expect ultimately a positive result from relic axion search experiments.
\end{abstract}

\maketitle

%%%%%%%%%%%%%%%%%%%%%%%%%%%%%%%%%%%%%%%%%%%%
%% MAINMATTER
%%%%%%%%%%%%%%%%%%%%%%%%%%%%%%%%%%%%%%%%%%%%

\section{Introduction}

The idea of the Standard Model extended by weak scale broken supersymmetry (SUSY) 
is extremely attractive in that it stabilizes the weak scale against quantum corrections,
and allows ultimately for an embedding in Grand Unified Theories (GUTs)\cite{dg}. GUTs
are also highly attractive in that they unify the disparate forces contained in the
SM into (usually) a single gauge group. SUSY GUTs receive some well-known indirect 
support from experiment in that the three gauge couplings-- when extrapolated up to
$Q\sim 2\times 10^{16}$ GeV via the MSSM RGEs\cite{drw}-- very nearly meet at a point!

The gauge group $SO(10)$ is especially compelling\cite{so10}. Not only does it unify the SM forces,
but it also unifies the SM {\it matter} of each generation 
into the 16-dimensional spinor representation. This unification only works if there is 
in addition to the SM superfields, a gauge singlet $\hat{N}^c_i$, for generations $i=1-3$, which
contains a right-hand neutrino state, as is required for see-saw neutrino masses\cite{seesaw}.
$SO(10)$ is naturally anomaly-free, thus explaining the otherwise ad-hoc anomaly cancellation
in the SM or in $SU(5)$. In addition, $SO(10)$ provides a basis for $R$-parity conservation, 
in that only {\it matter-matter-Higgs} couplings are allowed, while the $R$-violating
{\it matter-Higgs} or {\it matter-matter-matter} couplings are forbidden. If $SO(10)$ is broken
properly, the $R$-parity survives as an exact symmetry\cite{spmartin}. The simplest $SO(10)$ models
also allow for Higgs unification, since both $H_u$ and $H_d$ live in the 10 of $SO(10)$.
Finally, in the simplest models, we also expect $t-b-\tau$ third generation Yukawa coupling unification 
at $Q=M_{GUT}$. The above features have convinced many theorists that the main ideas behind 
$SO(10)$ SUSY GUTs are surely right (even while most or all explicit models in the
literature are likely wrong).

Here, we will assume (motivated by gauge coupling unification) that the MSSM, or MSSM plus
gauge singlets, is the correct effective field theory valid from $M_{weak}\sim 1$ TeV all the
way up to $M_{GUT}\sim 2\times 10^{16}$ GeV. We will also require that the third generation
$t-b-\tau$ Yukawa couplings should unify to reasonable ($\alt 5\%$) precision at $M_{GUT}$.

To test Yukawa unification, we scan over $SO(10)$-inspired SUSY parameter space:
\begin{equation}
m_{16},\ m_{10},\ M_D^2,\ m_{1/2},\ A_0,\ \tan\beta,\ sign(\mu ) .
\end{equation}
Here, $m_{16}$ is the common matter-scalar mass at $M_{GUT}$, $m_{10}$ the common Higgs mass, 
$m_{1/2}$ the common gaugino mass and $A_0$ the common trilinear soft term. 
$M_D^2$ parametrizes the Higgs multiplet splitting\cite{mur}, 
{\it i.e.} $m_{H_{u,d}}^2=m_{10}^2\mp 2M_D^2$
as is given by $D$-term mass contributions arising from the breaking of $SO(10)$.
We will examine two cases: the ``just-so'' Higgs splitting (HS) model, 
where only the Higgs scalars split, 
and the DR3 model, where full scalar $D$-term splitting is invoked, along with right-hand neutrino
contributions, and possible third generation scalar mass splitting ({\it i.e.} 
$m_{16}(3)\ne m_{16}(1,2)$ at $M_{GUT}$). 

We use the Isajet/Isasugra sparticle mass spectrum calculator\cite{isajet}. 
This includes full two-loop RGEs\cite{mv}, an  RG-improved one-loop effective potential calculation 
(evaluated at an optimized scale to account for leading two-loop effects) 
and full 1-loop sparticle mass corrections\cite{pbmz}. Especially important is including
the weak scale $t$, $b$ and $\tau$ self energy corrections when transitioning from MSSM to SM
effective theories; these depend on the entire superparticle mass spectrum, and are especially large 
for the $m_b$ correction at large $\tan\beta$\cite{hrs}.

Exhaustive scans over parameter space reveal that $t-b-\tau$ Yukawa unification only occurs
when the following conditions are met\cite{bf,bdr,us}:
\begin{itemize}
\item $A_0^2=2m_{10}^2=4m_{16}^2$, with
\item $m_{16}\sim 8-20$ TeV, 
\item $m_{1/2}$ very small $\sim 20-100$ GeV),
\item $\tan\beta\sim 50$
\item $m_D^2 >0$.
\end{itemize}
These conditions, derived earlier by Bagger {\it et al.}\cite{bfpz}, yield a 
{\it radiatively driven inverted scalar mass hierarchy} (RIMH). The physical 
sparticle mass spectrum is then given by
\begin{itemize}
\item first/second generation squarks and sleptons $\sim 8-20$ TeV,
\item third generation squarks, sleptons, $m_A$ and $\mu$: $\sim 1-2$ TeV,
\item light gauginos with $m_{\tg}\sim 300-500$ GeV, $m_{\tw_1}\sim 100-180$ GeV and
$m_{\tz_1}\sim 50-80$ GeV.
\end{itemize}
The heavy first/second generation squarks and sleptons can act to suppress
possible SUSY FCNC and $CP$ violating interactions, and proton decay.
The much lighter third generation scalars meet the needs for technical
naturalness.
Note that Yukawa-unified SUSY provides a viable realization of the ``effective SUSY'' scenario
put forth by Cohen, Kaplan and Nelson\cite{ckn}, while maintaining the MSSM as the correct 
effective theory all the way up to $M_{GUT}$.

\begin{figure}
  \includegraphics[height=.3\textheight]{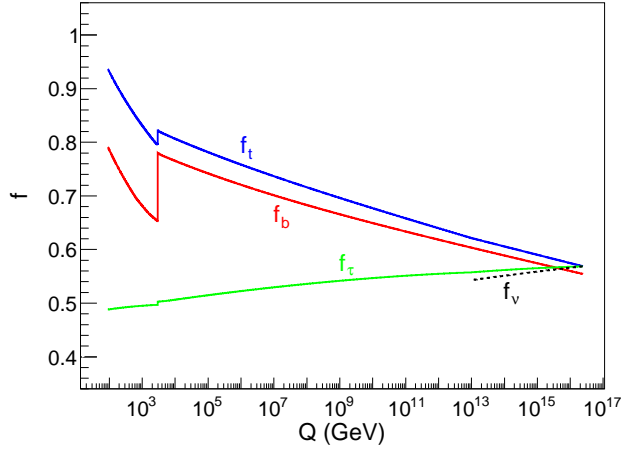}
  \caption{$t-b-\tau -\nu_{\tau}$ Yukawa unification in the DR3 model (from Ref. \cite{dr3}).}
\label{fig:yuk}
\end{figure}

The HS model is found to give many cases with {\it exact} $t-b-\tau$ unification\cite{bdr,us} 
(which is perhaps better than expected, given the theoretical uncertainties of the 
perturbative calculations). The DR3 model\cite{dr3}, 
with full $D$-term splitting, can give Yukawa unification 
down to the 2\% level, but only if neutrino Yukawa running is included down to $Q\sim 10^{13}$ GeV
(as suggested by neutrino mass difference measurements), and there is a small
first/third generation scalar splitting at $M_{GUT}$. An example case is shown in Fig. \ref{fig:yuk}.

\section{Mixed axion/axino cold dark matter}

If SUSY is broken in gravity mediation, then one expects the scalar masses to be directly related
to the gravitino mass. In this case, $m_{3/2}\sim 8-20$ TeV as well. This range of $m_{3/2}$
solves a major portion of the cosmological gravitino problem: such a heavy gravitino will have a lifetime
less than 1 second\cite{moroi}, so that it decays just before the onset of BBN. Thus, this model should be 
BBN safe, and should allow for a re-heat temperature $T_R\sim 10^6-10^9$ GeV.
While this $T_R$ value is not enough to sustain thermal leptogenesis as a baryogenesis
mechanism, 
it is enough to sustain non-thermal leptogenesis, 
wherein right hand neutrinos are produced via
inflaton decay, or Affleck-Dine leptogenesis.

The above mass spectrum predicted by Yukawa-unified SUSY has many desirable features.
However, if we calculate the thermally produced relic abundance of neutralinos
(we use IsaReD\cite{isared}), we find $\Omega_{\tz_1}h^2\sim 10^2-10^4$, {\it i.e.}
3-5 orders of magnitude higher than the measured value\cite{relic}!

At this point, we have totally neglected (at our peril!) the strong $CP$ problem. If we invoke
the PQWW\cite{pqww} solution to the strong CP problem with an ``invisible'' axion\cite{invax},
then we must include the axion/axino supermultiplet in the theory. The QCD axion
has mass $m_a\sim 10^{-6}-10^{-3}$ eV according to astrophysical/cosmological constraints.
The axino $\ta$ is $R$-parity odd, and can serve as the LSP\cite{rtw}. Its mass is relatively
unconstrained, and can span the keV$\to$ GeV range. If $m_{\ta}\sim$ MeV range, then
$\tz_1\to\ta\gamma$ with a lifetime typically less than 1 second (BBN safe). Each thermally produced
neutralino will decay to exactly one axino, and the (non-thermally produced) axino abundance
will be $\Omega_{\ta}^{NTP}h^2=\frac{m_{\ta}}{m_{\tz_1}}\Omega_{\tz_1}h^2$\cite{ckkr}: the ratio of masses
yields a factor $\sim 10^{-3}-10^{-5}$, and can completely wipe out the neutralino
overabundance. These decay-produced MeV-scale axinos would likely constitute 
{\it warm} dark matter\cite{jlm}.

The axinos can also be produced thermally via scattering off quarks and gluons early on 
in the cosmic soup. The abundance depends on the axino mass, the PQ breaking scale $f_a$ and
the re-heat temperature $T_R$ after inflation\cite{ckkr,bsteff}. The thermally produced axinos will
constitute cold dark matter so long as $m_{\ta}\agt 0.1$ MeV.

A third dark matter component comes as usual from vacuum misalignment production
of cold axions, as shown by Sikivie and others\cite{as}. This contribution $\Omega_ah^2$
depends on the PQ scale $f_a$ (or alternatively the axion mass) 
and the initial mis-alignment angle $\theta_i$. 

The total dark matter abundance in the PQMSSM model comes from three components:
\begin{equation}
\Omega_{DM}h^2=\Omega_{\ta}^{NTP}h^2+\Omega_{\ta}^{TP}h^2+\Omega_a h^2 .
\end{equation}
We examined Yukawa-unified SUSY models with both low and high values of $f_a$ and $m_{\ta}$\cite{bhkss}.
By enforcing $\Omega_{DM}h^2=0.1$ as measured by WMAP, we can extract the required 
re-heat temperature. The values of $T_R$ can range easily between $10^6-10^8$ GeV, thus
being in the range required by the gravitino problem, and also allowing for
non-thermal or Affleck-Dine leptogenesis!

\section{Consequences for Tevatron, LHC and ADMX}

The Yukawa-unified SUSY model predicts a rather light gluino with $m_{\tg}\sim 300-500$ GeV.
In standard SUSY models with gaugino mass unification, the LEP2 limit on $m_{\tw_1}>103.5$ GeV
implies $m_{\tg}\agt 420$ GeV, which is somewhat beyond Tevatron reach. However, in Yukawa-unified SUSY,
the huge $A_0\sim 20$ TeV parameter feeds into gaugino mass evolution via two-loop RGEs
to suppress the gap between the $SU(2)$ and $SU(3)$ gaugino masses $M_2$ and $M_3$. Thus, in Yukawa-unified
SUSY, $m_{\tg}$ can be as low as $\sim 300$ GeV while respecting the LEP2 chargino limit.
Also, the huge first generation squark masses actually {\it suppress} negative interference in the
$q\bar{q}\to\tg\tg$ production cross section, thus raising the $\tg\tg$ production cross section at the
Tevatron by factors of 3-10 beyond standard calculations! Finally, the $\tg$ decays nearly 100\%
of the time via 3-body modes into $b$-quarks. Thus, four or more $b$-jets are expected in each final state.
By requiring $n_b\ge 2$ or even $n_b\ge 3$, SM backgrounds are highly suppressed. Detailed calculations find
a 5 fb$^{-1}$ reach of Tevatron to $m_{\tg}\sim 400$ GeV\cite{tev}. This probes the {\it most favored} portion
of Yukawa-unified parameter space, since Yukawa-unification worsens as $m_{\tg}$ increases. 
{\bf This search is strongly recommended for CDF and D0!}

\begin{figure}
  \includegraphics[height=.3\textheight]{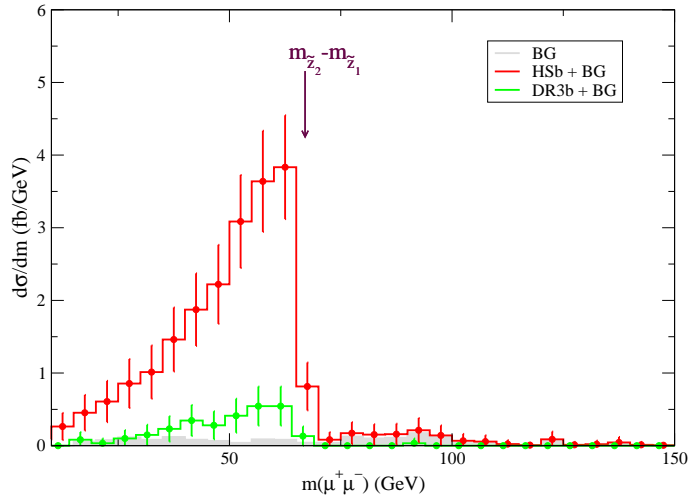}
  \caption{Opposite-sign dimuon invariant mass distribution after cuts at LHC for 
Yukawa unified SUSy in the DR3 and HS models\cite{lhc7}.}
\label{fig:mll}
\end{figure}

At the $\sqrt{s}=7$ TeV LHC, the $pp\to\tg\tg X$ cross section ranges between 2000-6000 fb, and occurs
mainly via $gg$ annihilation. The events should again be characterized by high $b$-jet multiplicity.
In addition, the decay $\tg\to b\bar{b}\tz_2$ followed by $\tz_2\to\tz_1\ell^+\ell^-$ can occur
at a large rate. In this case, the $m(\ell^+\ell^-)$ distribution for SF/OS dileptons should have a
characteristic mass bump followed by an edge at $m_{\tz_2}-m_{\tz_1}$: see Fig. \ref{fig:mll}. 
Since this mass gap is typically bounded
by about 90 GeV, the bump/edge should sit between the $\gamma$ and $Z$ peaks, and should be easily
visible even with very low integrated luminosity $\sim 0.1$ fb$^{-1}$. An estimate of the $\sqrt{s}=7$ TeV LHC reach
with just 0.1 fb$^{-1}$ of integrated luminosity, and not using $\eslt$ cuts, is to $m_{\tg}\sim 400$ GeV. 
With 1 fb$^{-1}$ of integrated luminosity, which is now anticipated in LHC run 1, the reach using $\eslt$ cuts
is to $m_{\tg}\sim 630$ GeV. Thus, it is expected that LHC will be able to either discover or rule out
Yukawa-unified SUSY during its first run at $\sqrt{s}=7$ TeV!
{\bf This search is strongly recommended for Atlas and CMS!}

Finally, since a large value of $f_a$ is favored cosmologically, we would expect the axion/axino
abundance to be mainly mis-alignment produced axions. Thus, we can anticipate a possible axion discovery
by ADMX\cite{admx} in the years to come, if Yukawa-unified SUSY is correct.

\section{Conclusions}

The Yukawa-unified SUSY scenario invokes IMO the four greatest ideas in physics beyond the SM:
grand unification, supersymmetry, see-saw neutrino masses and the PQWW axion solution to the strong CP problem.
Yukawa-unified SUSY with mixed axion/axino CDM also solves several cosmological problems, and is
consistent with low energy realizations of the fifth greatest idea: string theory\cite{raby}.
The immediate prediction is a rather light gluino with $m_{\tg}\sim 300-500$ GeV which decays via three-body
modes into mainly $b$-quarks: it should be observable in the next year or two via Tevatron and LHC run 1 data.
A positive signal would also be likely at the ADMX experiment.

%%%%%%%%%%%%%%%%%%%%%%%%%%%%%%%%%%%%%%%%%%%%

%%%%%%%%%%%%%%%%%%%%%%%%%%%%%%%%%%%%%%%%%%%%%%%%
%% BACKMATTER
%%%%%%%%%%%%%%%%%%%%%%%%%%%%%%%%%%%%%%%%%%%%%%%%

\begin{theacknowledgments}
I thank D. Auto, C. Balazs, A. Belyaev, J. Ferrandis, 
S. Kraml, A. Lessa, S. Sekmen, H. Summy, and X. Tata 
for a fruitful collaboration,
and I wish Pierre a happy 60th!
\end{theacknowledgments}

\bibliographystyle{aipproc}   % if natbib is available

%%%%%%%%%%%%%%%%%%%%%%%%%%%%%%%%%%%%%%%%%%%
%% The following lines show an example how to produce a bibliography
%% without the help of the BibTeX program. This could be used instead
%% of the above.
%%%%%%%%%%%%%%%%%%%%%%%%%%%%%%%%%%%%%%%%%%%

\end{document}